\documentclass[acus]{JAC2000}

\usepackage{graphicx}
\usepackage{epsfig}
\usepackage{rotating} 
\usepackage{multicol}

\usepackage{relsize}
\def\babar{\mbox{\slshape B\kern-0.1em{\smaller A}\kern-0.1em
    B\kern-0.1em{\smaller A\kern-0.2em R}}}

\def\en         {\ensuremath{e^-}}      
\def\ep         {\ensuremath{e^+}}

\def\mumu       {\ensuremath{\mu^+\mu^-}}

\def\pipi  {\ensuremath{\pi^+\pi^-}}

\def\Kbar  {\ensuremath{\kern 0.2em\overline{\kern -0.2em K}}}

\def\Kstarb   {\ensuremath{{\Kbar}\kern0.03em {\raise.18ex\hbox{$^*$}}}}

\def\Kzb   {\ensuremath{{\Kbar}\kern0.03em {\raise.1ex\hbox{$^0$}}}}
\def\KzKzb {\ensuremath{K^0 {\kern -0.16em \Kzb}}}

\def\Dbar  {\ensuremath{\kern 0.2em\overline{\kern -0.2em D}}}
\def\Dzb   {\ensuremath{{\Dbar}\kern0.03em {\raise.3ex\hbox{$^0$}}}}
\def\DzDzb {\ensuremath{D^0 {\kern -0.16em \Dzb}}}

\def\Dstarb   {\ensuremath{{\Dbar}\kern0.03em {\raise.18ex\hbox{$^*$}}}}

\def\Bbar  {\ensuremath{\kern 0.18em\overline{\kern -0.18em B}}}
\def\Bzb   {\ensuremath{{\Bbar}\kern0.03em {\raise.3ex\hbox{$^0$}}}}

\def\BzBzb {\ensuremath{B^0 {\kern -0.16em \Bzb}}}

\mathchardef\Upsilon="7107
\def\Y#1S{\ensuremath{\Upsilon{(#1S)}}}

\mathchardef\Delta="7101
\mathchardef\Xi="7104
\mathchardef\Lambda="7103
\mathchardef\Sigma="7106
\mathchardef\Omega="710A
\def\Deltabar   {\ensuremath{\kern 0.25em\overline{\kern -0.25em \Delta}}{}}
\def\Lbar {\ensuremath{\kern 0.2em\overline{\kern -0.2em\Lambda\kern 0.05em}\kern-0.05em}{}}
\def\Sigbar{\ensuremath{\kern 0.2em\overline{\kern -0.2em \Sigma}}{}}
\def\Xibar{\ensuremath{\kern 0.2em\overline{\kern -0.2em \Xi}}{}}
\def\Obar{\ensuremath{\kern 0.2em\overline{\kern -0.2em \Omega}}{}}
\def\Nbar{\ensuremath{\kern 0.2em\overline{\kern -0.2em N}}{}}

\def\ev   {\ensuremath{\rm \,e\kern -0.08em V}}
\def\kev  {\ensuremath{\rm \,ke\kern -0.08em V}} 
\def\mev  {\ensuremath{\rm \,Me\kern -0.08em V}} 
\def\gev  {\ensuremath{\rm \,Ge\kern -0.08em V}} 
\def\gevc {\ensuremath{{\rm \,Ge\kern -0.08em V\!/}c}} 
\def\tev  {\ensuremath{\rm \,Te\kern -0.08em V}}
\def\mevc {\ensuremath{{\rm \,Me\kern -0.08em V\!/}c}} 
\def\gevcc{\ensuremath{{\rm \,Ge\kern -0.08em V\!/}c^2}} 
\def\mevcc{\ensuremath{{\rm \,Me\kern -0.08em V\!/}c^2}}

\def\mus  {\ensuremath{\rm \,\mus}}

\def\mus        {\ensuremath{\,\mu{\rm s}}}    
\renewcommand{\bar}[1]{\overline{#1}}

\def\gsim{{~\raise.15em\hbox{$>$}\kern-.85em
          \lower.35em\hbox{$\sim$}~}}
\def\lsim{{~\raise.15em\hbox{$<$}\kern-.85em
          \lower.35em\hbox{$\sim$}~}}

\def\to                 {\ensuremath{\rightarrow}}

\def\pep2{PEP-II}

\def\chic1{\ensuremath{\chi_{c1}}}
\def\chic2{\ensuremath{\chi_{c2}}}
\def\chic3{\ensuremath{\chi_{c3}}}

\newcommand{\eqref}[1]{Eq.~(\ref{eq:#1})}

\def\jetset74   {\mbox{\tt Jetset \hspace{-0.5em}7.\hspace{-0.2em}4}}

\def\kk      {\ensuremath{K^+K^-}}

\begin{document}

\title{\flushright{PSN T03}\\[15pt] \centering 
STUDY OF $\ep\en$ COLLISIONS IN THE 1.5-3 GEV C.M. ENERGY REGION
USING ISR AT \babar
\thanks{Invited talk at the workshop ``$\ep\en$ Physics at
Intermediate Energies'', SLAC, Stanford, April 30 - May 2,
2001. The author is grateful to SLAC for support.
}
}
\author{E. P. Solodov, Budker Institute of
Nuclear Physics, Novosibirsk, Russia\\
For the \babar~Collaboration}
 
\maketitle

\begin{abstract}

A preliminary analysis of low-energy $\ep\en$ collision data produced
via initial state radiation (ISR)
has been performed using 22 $fb^{-1}$ of \babar ~data. The selection of
data samples corresponding to the
$\mumu$, $\pipi$, $\kk$, $p \bar p$, $\kk\pi^0$, 3$\pi$, 4$\pi$,
5$\pi$, 6$\pi$, 7$\pi$ final states
accompanied by the emitted ISR hard photon
has been demonstrated.
The invariant mass of the hadronic final state defines
the effective collision c.m.
energy, and so \babar~ ISR data can be compared to the relevant
direct $\ep\en$ measurements. The resulting distributions
are already competitive with DCI and ADONE
data in the 1.4-3.0 GeV energy range. In particular, they do not suffer
from the relative normalization uncertainties observed
for certain reactions when results from different
experiments are combined.
Eventually, such data may be used to measure the energy dependence
of R, the ratio of the $\ep\en\to hadrons$ and
 $\ep\en\to\mumu$ cross sections, in the low-energy regime where precise
measurements will have an impact on the interpretation of the new
$(g-2)_{\mu}$ measurements.
\end{abstract}

\section{Introduction}\label{intro}

The possibility of using the initial state radiation (ISR) of hard
photons at B-factories to study hadronic final state production
in $\ep\en$ collisions at lower c.m. energies
has been discussed previously~\cite{ivanch, kuehn}.
Preliminary studies of some particular ISR
processes have been performed with \babar ~data
~\cite{vuco, Lou}.
This paper reports preliminary results from
a relatively simple pilot analysis of
exclusive hadronic final states accompanied by a hard
(1-9 GeV) photon assumed to result from ISR. 
Events corresponding to
$\ep\en\to\mumu\gamma$ are selected also, since these enable the
normalization of the hadronic cross section measurements.
ISR photons are produced at all angles relative to
the collision axis, and it has been shown ~\cite{ivanch} that the \babar~
acceptance for such photons is around 10-15 $\%$.
The analysis samples are selected from
22 $fb^{-1}$ of \babar ~ $\Upsilon(4S)$ and continuum data collected
in 1999-2000. 

 
The ISR cross section for a particular final state $f$, with 
$\ep\en$ cross section $\sigma_{f}(s)$, is obtained,
to first order, from:\\

$ \frac{d\sigma (s,x)}{dx} = W(s,x)\cdot \sigma_{f}(s(1-x))$, \\ 

where ~~$x=\frac{2E_{\gamma}}{\sqrt{s}}$; $E_{\gamma}$ is the
energy of the ISR photon in the nominal c.m. frame, and $\sqrt{s}$ is 
the nominal c.m. energy.
 The function \\
 
$
 W(s,x) = \beta\cdot( (1+\delta)\cdot x^{(\beta-1)}-1+\frac{x}{2} )
$\\

describes the energy spectrum of the ISR photons; \\

$
 \beta = \frac{2\alpha}{\pi x}\cdot (2ln \frac{\sqrt{s}}{m_{e}}-1) ,
$\\

and $\delta$ takes into account vertex and self-energy corrections.
 At the ~ $\Upsilon(4S)$ energy, $\beta$ = 0.088 and $\delta$ = 0.067.

 
For a hadronic final state, $f$, the normalized cross section at c.m.
energy squared $s'$, $\sigma_{f}(s')$, is obtained by relating
the observed number of events in an interval $ds'$ centred at $s'$,
$dN_{f\gamma}$, to the corresponding number of radiative dimuon events,
$dN_{\mu\mu\gamma}$, by means of \\

$\sigma_{f}(s') =
\frac
{dN_{f\gamma}\cdot\epsilon_{\mu\mu}\cdot(1+\delta_{rad}^{\mu\mu})}
{dN_{\mu\mu\gamma}\cdot\epsilon_{f}\cdot(1+\delta_{rad}^{f})}
\cdot \sigma_{e^+e^-\to\mu+\mu-}(s') $, \\

where $s' = s(1-x) $;
$\epsilon_{\mu\mu}$ and $\epsilon_{f}$ are detection
efficiencies, and $1+\delta_{rad}^{\mu\mu}$, $1+\delta_{rad}^{f}$ are
final state radiative correction factors.
The radiative corrections to the initial state,
acceptance for the ISR photon, and virtual photon properties are
the same for $\mumu$ and $f$, and cancel in the ratio.

\section{$\mumu\gamma$ final state and effective luminosity}

\babar~ provides excellent particle identification (PID) information,
and this permits the selection of two-prong events
containing a hard photon, for which at least one of the charged
tracks is well-identified as a muon.

\begin{figure}[hbt]
{
\epsfxsize=9.0cm
\epsfysize=8.5cm
\hspace{-0.8cm}
\epsfbox{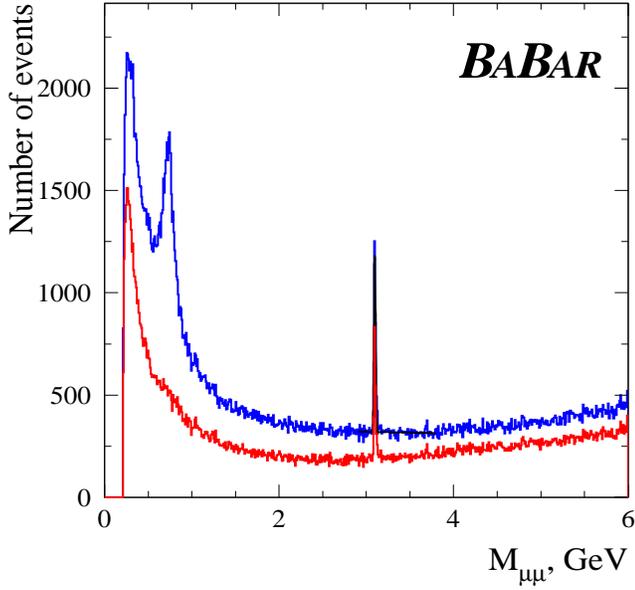}
\vspace{-0.9cm}
\caption
{The $\mumu$ invariant mass distribution for two-prong events with a
hard photon, when one track is identified as a muon (upper histogram),
and when both tracks are muon-identified(lower histogram).}
}
\label{mumu1}
\end{figure}

\begin{figure}[hbt]
{
\epsfxsize=9.0cm
\epsfysize=8.5cm
\hspace{-0.8cm}
\epsfbox{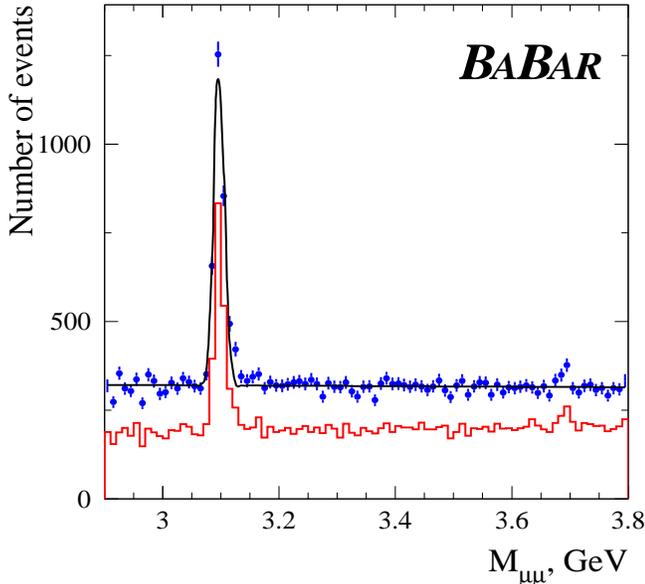}

\vspace{-1.0cm}
\caption
{ Fig. 1 in the $J/\psi$ region.}
}
\label{mumu2}
\end{figure}

Fig.~1 shows the $\mumu$ invariant mass distributions for these
events when at least one track is muon-identified (upper histogram), 
and when both tracks are muon-identified (lower histogram).
In the upper distribution, the peak
at 0.8 GeV is due to $\ep\en\to\rho\gamma\to\pipi\gamma$;
it disappears almost entirely when the
second track also is identified as a muon. A sharp peak due to
$J/\psi$ decay to $\mumu$ is present in both distributions.

Fig. ~\ref{mumu2} is a close-up view of fig. 1 in the
$J/\psi$ region. A fit using a Gaussian line-shape and
linear background gives $\sim$2000 events from $J/\psi$ decay,
and there is a small signal due to $\psi(2S)$ decay at $\sim$3.7 GeV.

\begin{figure}[tbh]
{
\epsfxsize=9.0cm
\epsfysize=8.5cm
\hspace{-0.7cm}
\epsfbox{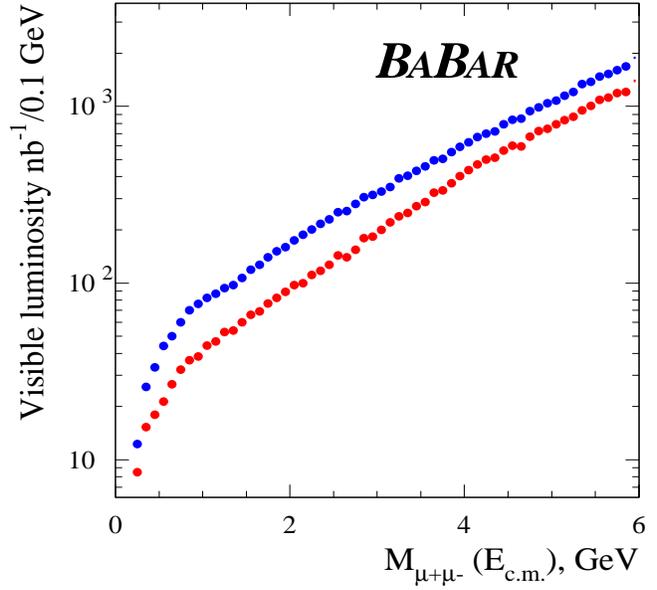}
}
\vspace{-1.0cm}
\caption
{The visible (detected) integrated luminosity per 0.1 GeV
 for ISR at \babar~ with no
 correction for acceptance and final state radiation
 when at least one track is muon-identified (upper distribution),
and when both tracks are muon-identified (lower distribution).
}
\label{mumu_lum}
\end{figure}

The $\mumu\gamma$ events can be used to normalize
ISR production of hadronic final states. The
invariant mass of the muon pair
defines the effective collision energy, i.e. the c.m. energy of
the virtual photon. The energy dependence
of the integrated luminosity, $dL$,
for the interval $dE_{\gamma^*}$ centred
at virtual photon energy $E_{\gamma^*}$ is then obtained from\\
\\
$dL(E_{\gamma^*})=\frac
{  dN_{\mu\mu\gamma}(E_{\gamma^*}) }
{
\epsilon_{\mu\mu}\cdot(1+\delta_{rad}^{\mu\mu})
\cdot\sigma_{\ep\en\to\mumu}(E_{\gamma^*}) 
}
$,~~~ $E_{\gamma^*}=m_{inv}^{\mu\mu}$\\

where $dN_{\mu\mu\gamma}$ is the number of dimuon events observed in
this interval.
Fig.~\ref{mumu_lum} shows the energy dependence of
the visible (i.e. not corrected for
acceptance and final state radiation) luminosity integrated over 
0.1 GeV intervals.
The upper distribution is obtained when at least one final state track
is identified as a muon, while the lower requires both tracks to be
muon-identified.

An advantage deriving from the use of ISR is that the
entire range of effective collision energy is scanned in one
experiment. This avoids the relative normalization uncertainties which 
can arise when data from different experiments are combined.
The present \babar~ data are equivalent to an $\ep\en$ machine scan in
0.1 GeV steps with a luminosity integral per point varying
from ~100 nb$^{-1}$ at 1 GeV to ~1 pb$^{-1}$ at 5 GeV c.m.
energy.
 A disadvantage is that invariant mass resolution limits
the width of the narrowest structure which can be measured via
ISR production.
The resolution can be monitored directly using
the width of the $J/\psi$ signal; by using a kinematic fit, a value
of $\sim$8 MeV can be achieved for $\mumu$ events.

\begin{figure}[tbh]
{
\hspace{-1.0cm}
\epsfxsize=9.3cm
\epsfysize=8.9cm
\epsfbox{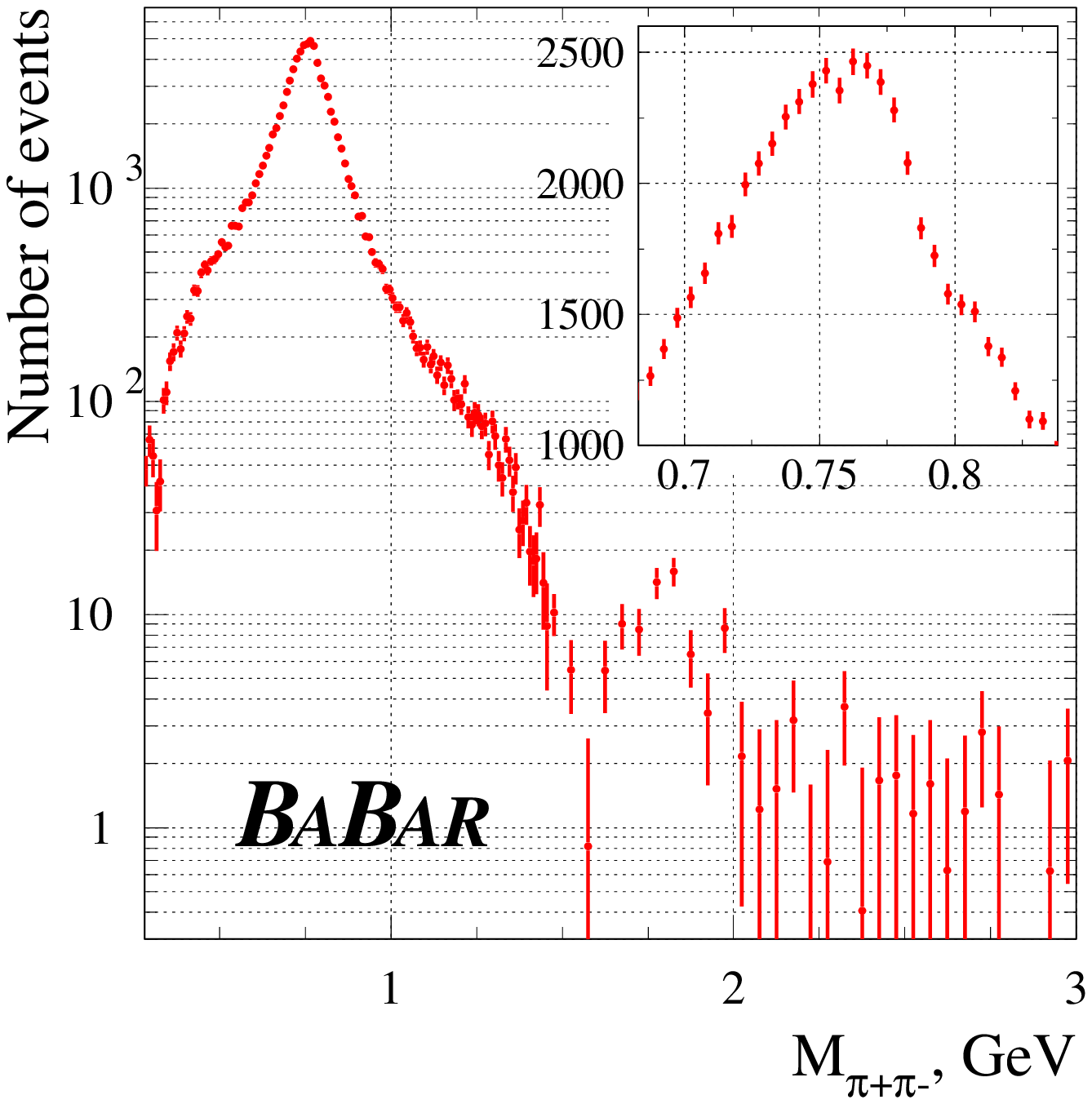}
\hspace{-1.2cm}
\epsfxsize=9.2cm
\epsfysize=7.5cm
\epsfbox{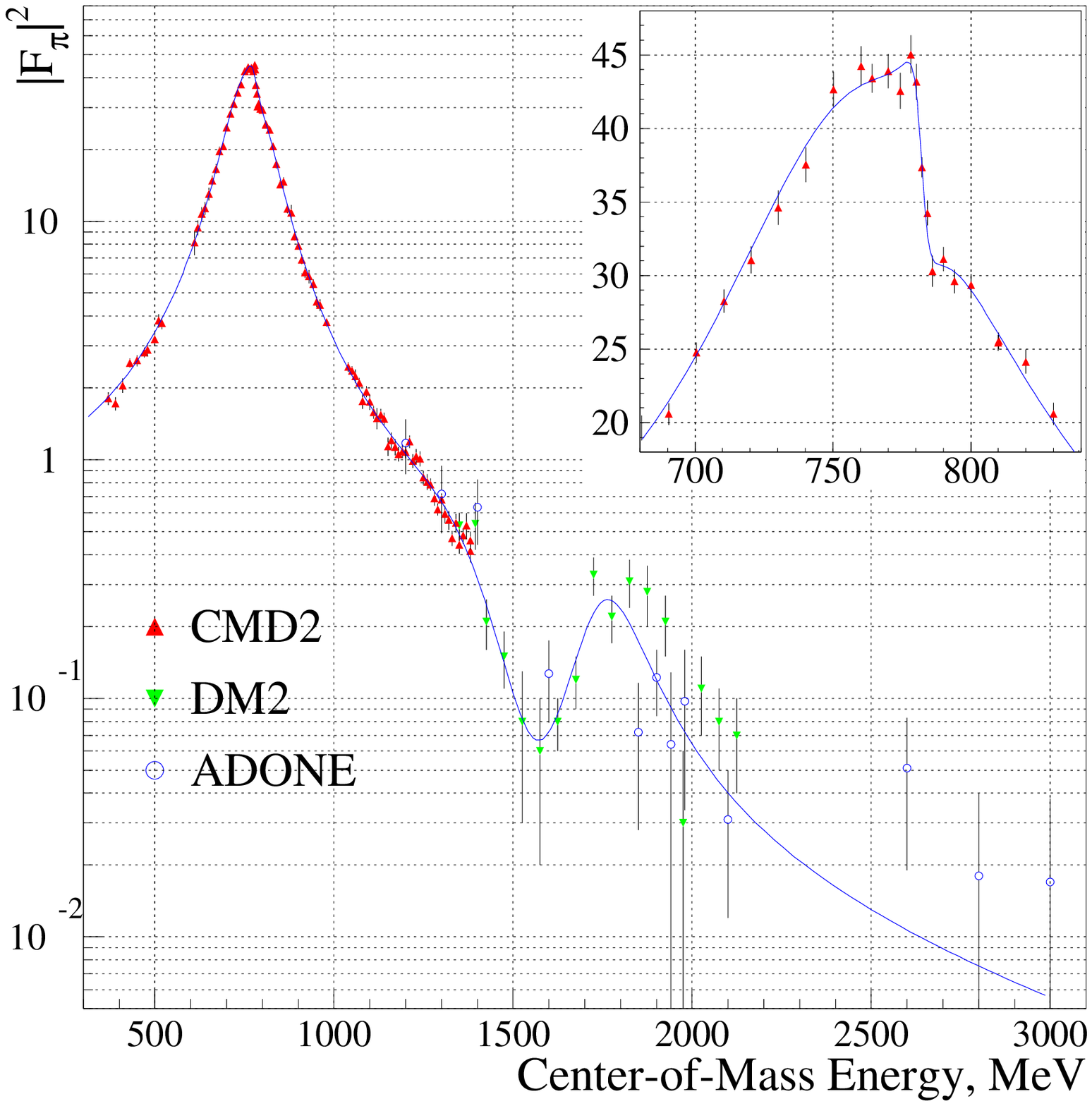}
}
\vspace{-0.5cm}
\caption
{ The uncorrected $\pipi$ invariant mass distribution for $\pipi\gamma$
 events from  \babar~ (top), compared to pion form factor
 measurements from VEPP-2M and DCI (bottom).
}
\label{pipi_invm}
\end{figure}

\section{Two hadron final states}
\subsection{$\pipi\gamma$ selection}\label{pipig}

The general selection criteria are the same for $\pipi\gamma$
and for $\mumu\gamma$ events. However, $\pipi$ selection
requires that neither final state charged track be identified
as a muon, kaon or proton (charge conjugation is implied throughout this
paper, whenever relevant). The $\rho$ peak dominates the resulting
dipion invariant mass distribution, but for mass greater than
2.0-2.5 GeV, dimuon feedthrough is the main contribution to the
spectrum.

\begin{figure}[b]
\vspace{-0.4cm}
{
\hspace{-0.5cm}
\epsfxsize=9.0cm
\epsfysize=8.5cm
\epsfbox{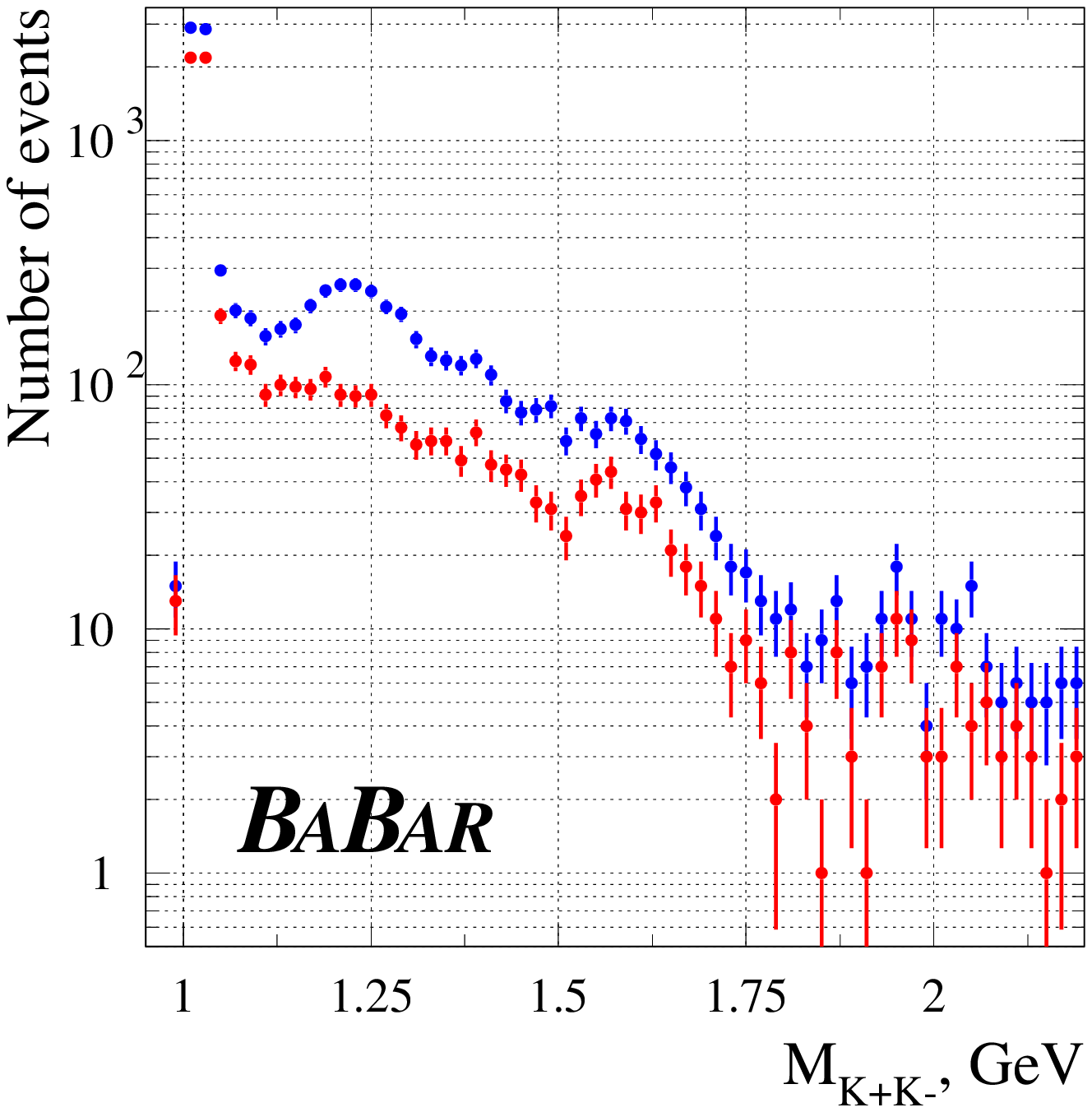}
\epsfxsize=8.5cm
\epsfysize=8.5cm
\epsfbox{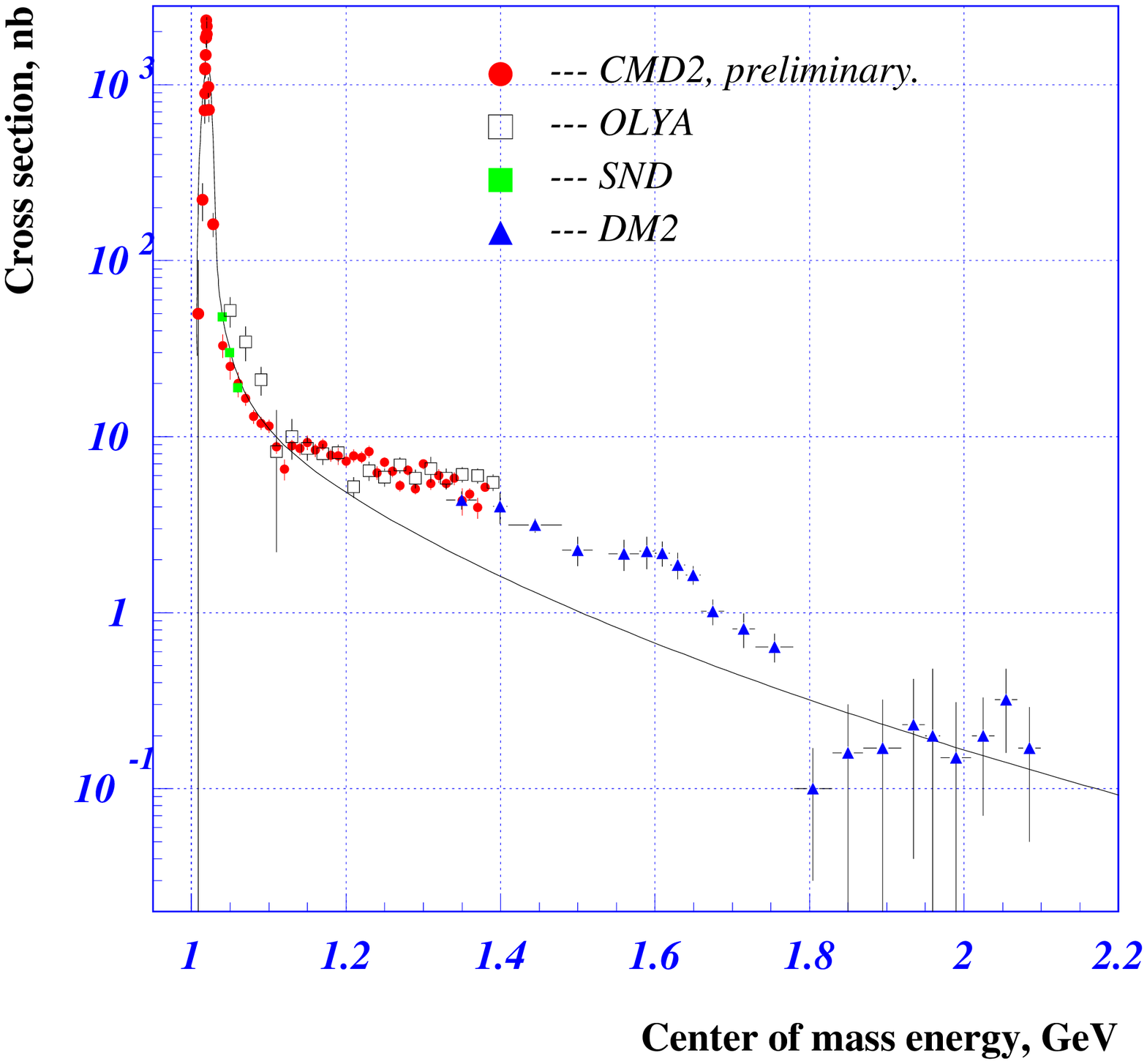}
}
\vspace{-0.9cm}
\caption
{ The uncorrected $\kk$ invariant mass distribution for $\kk\gamma$
 events from  \babar~ (top), compared to $\kk$ cross section
 measurements from VEPP-2M and DCI (bottom).
}
\label{kk_invm}
\end{figure}

 Assuming the dipion contribution above 3.5 GeV to be
negligible compared to that from dimuon background,
the mass distribution of fig.~\ref{mumu1} can be normalized
to this region and used in a background subtraction over the
entire mass range.

The resulting distribution is compared in
fig.~\ref{pipi_invm} (top) to data
on the pion form factor (bottom)~\cite{cmd2_pipi}. The
agreement is very encouraging. The insets show the
$\rho-\omega$ interference region. The form factor data show clear
evidence of e.m. mixing, and although there is evidence of similar
behavior in the data from \babar~, it is less sharply defined as a
consequence of the poorer mass resolution (~13 MeV at the $\rho$ meson
peak). Nevertheless, the effect appears to be present.

\subsection{$\kk\gamma$ selection}\label{KKg}

The requirement that at least one final state track be identified as a
kaon, and that neither be identified as a muon, pion or proton, leads to
a rather clean sample of $\kk\gamma$ events.
The resulting $\kk$ invariant mass distribution is shown in
fig.~\ref{kk_invm} (top) in comparison to the corresponding cross
section measurements from VEPP-2M and DCI (bottom). Both distributions
show a clear $\phi$ peak. The upper \babar~ data points are obtained by
requiring that at least one kaon be identified, while the lower
require that both be identified. The structure at ~1.2 GeV in the upper
distribution is due to feedthrough from the $\rho$; it is almost
entirely suppressed in the lower distribution, indicating the
effectiveness of the kaon identification procedure. The lower spectrum agrees
quite well with the cross section data (bottom), and it will be of
interest to learn whether or not the peaks in the \babar~ data at
~1.6 GeV and ~2.0 GeV, which are less clear in the cross section
data, are real.

\begin{figure}[hp]
\hspace{-0.7cm}
{
\epsfxsize=9.0cm
\epsfysize=8.0cm
\epsfbox{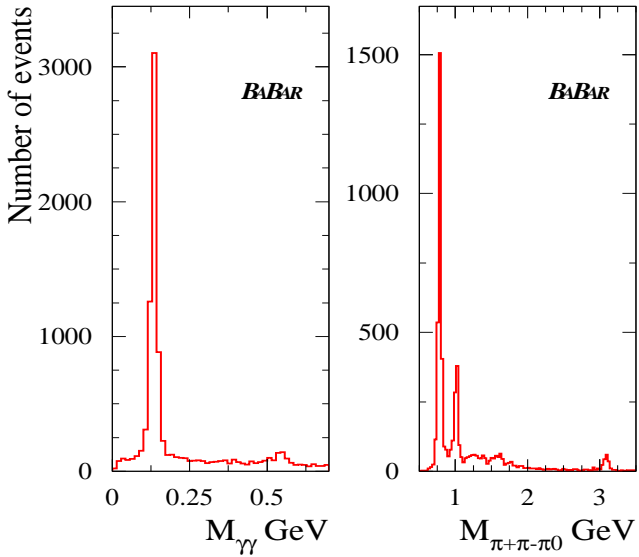}
}
\vspace{-0.9cm}
\caption
{ The $\gamma\gamma$ invariant mass distribution for $\pipi\gamma\gamma$
events (left); the combinations in the $\pi^0$ region were used to
construct the three pion invariant mass distribution shown in the
right-hand figure.
}
\label{gg_3pi_invm}
\end{figure}

\section{$\pi^+\pi^-\pi^0$ and $\pi^+\pi^-\eta$ final states }

For events with several photons, the most
energetic with energy greater than 1.0 GeV is interpreted as an ISR
candidate. The remaining photons having energy greater than 100 MeV
are then used in a $\pi^0$ and/or $\eta$ search.

 Two-prong events, with both tracks identified as pions, were required
to have three photons in addition to the ISR candidate.
The invariant mass distribution for the two softest photons from such
events is shown
in fig.~\ref{gg_3pi_invm}(left). Peaks are observed at the
$\pi^0$ and $\eta$ mass positions. Defining a $\pi^0$ candidate by
$|m_{\gamma\gamma}-m_{\pi^0}|<40$ MeV, the three pion
invariant mass distribution is as shown in the right-hand plot.
Clear peaks due to the $\omega$, $\phi$ and $J/\psi$ mesons are
observed, indicating that data samples corresponding to
the ISR production of these states with subsequent three-pion decay
can be readily selected.

\section{Four pion final states}
\subsection{$\pi^+\pi^-\pi^+\pi^-$ final state }

\begin{figure}[tbh]
\vspace{-0.2cm}
\centerline
{
\epsfxsize=8.5cm
\epsfysize=8.0cm
\epsfbox{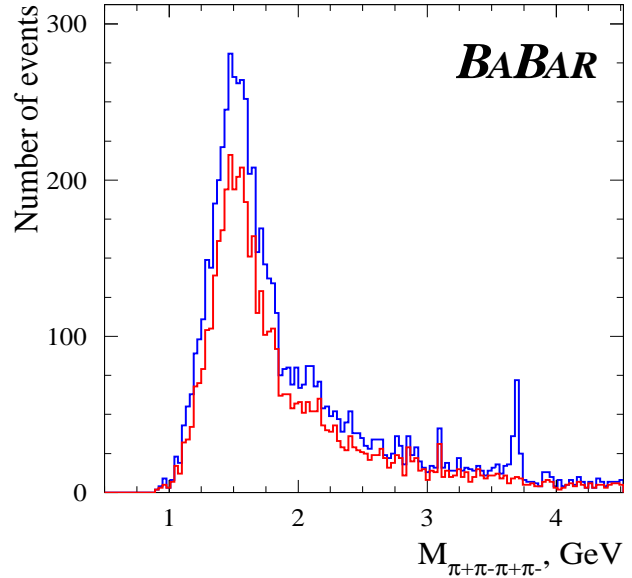}
}
\vspace{-0.3cm}
\caption{The invariant mass distribution for
four-prong ISR events from \babar~ when
 no track is identified as a kaon
 or proton (upper histogram), and when
 in addition no track is identified as a muon
(lower histogram).
}
\label{4pi_babar}
\end{figure}

A relatively clean sample of four-pion candidate events is selected
by requiring that, in addition to the ISR candidate photon,
there be four charged tracks, none of which is
identified as a kaon or proton.
Fig.~\ref{4pi_babar} shows the mass distribution
for the four charged tracks from such events (upper histogram). There is
a narrow peak at the $\psi(2S)$ due to the decay
$\psi(2S)\to\pipi J/\psi$ with $J/\psi\to\mumu$; this is
removed by the additional requirement that no track be identified
as a muon (lower histogram).

\begin{figure}[tbh]
\centerline
{
\epsfxsize=9.0cm
\epsfysize=8.5cm
\epsfbox{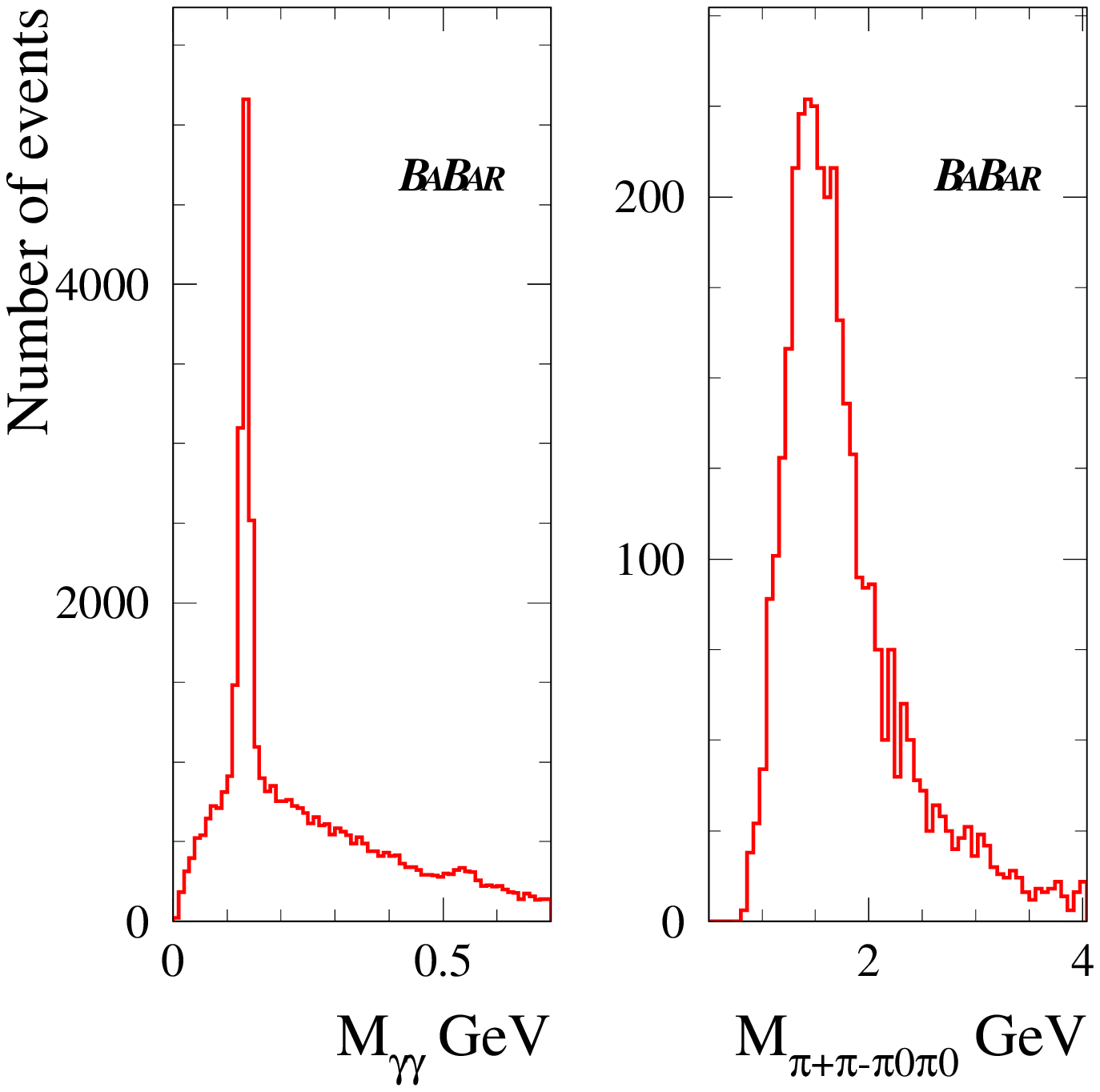}
}
\vspace{-0.3cm}
\caption{The two photon invariant mass distribution from 
$\pipi 4\gamma$ ISR events (left) and the
four pion invariant mass distribution obtained when two distinct photon
combinations are in the $\pi^0$ mass region.
}
\label{2pi2pi0_babar}
\end{figure}

\begin{figure}[tbh]
{
\epsfxsize=9.0cm
\epsfysize=8.0cm
\epsfbox{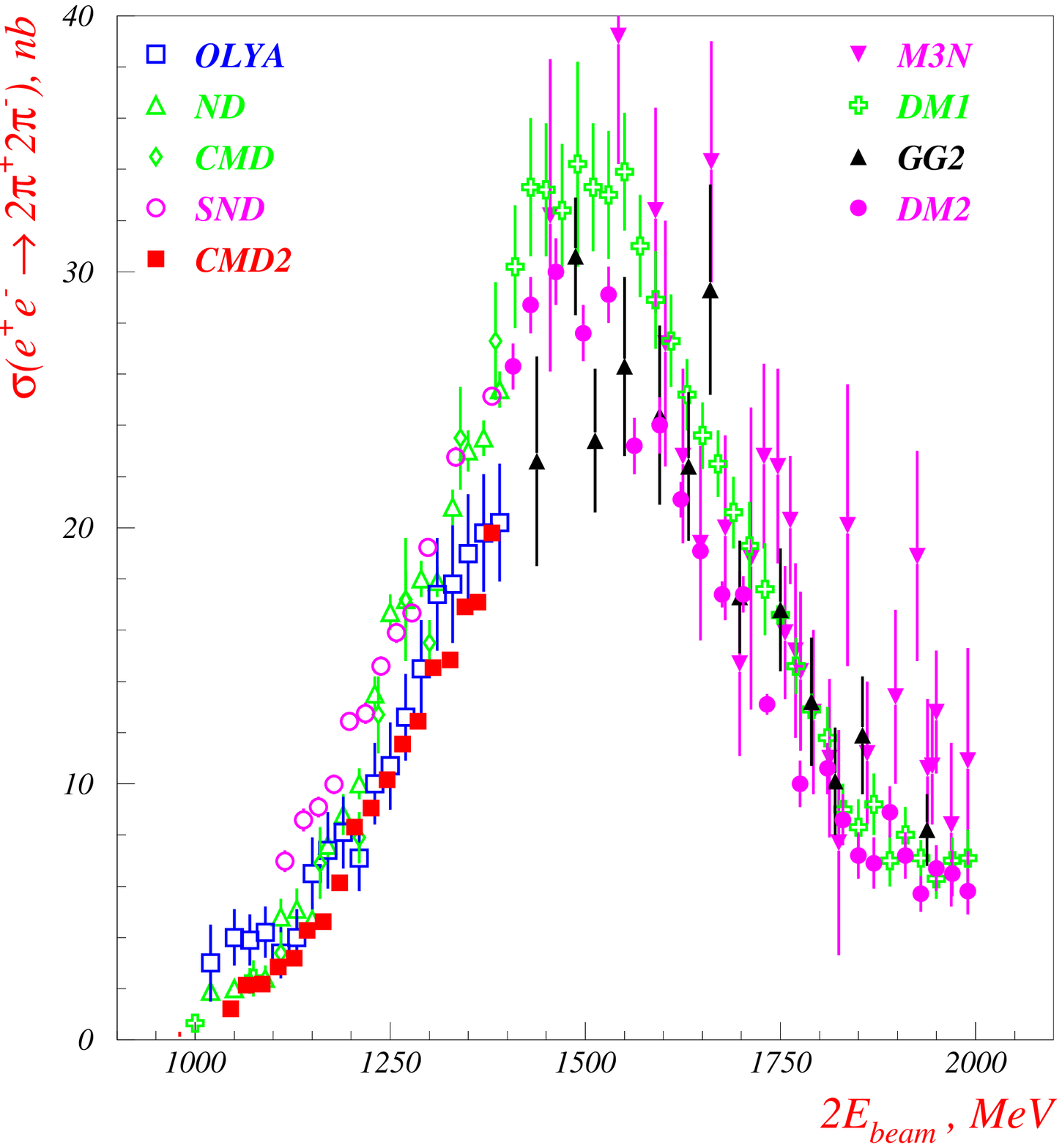}
\epsfxsize=8.3cm
\epsfysize=7.8cm
\epsfbox{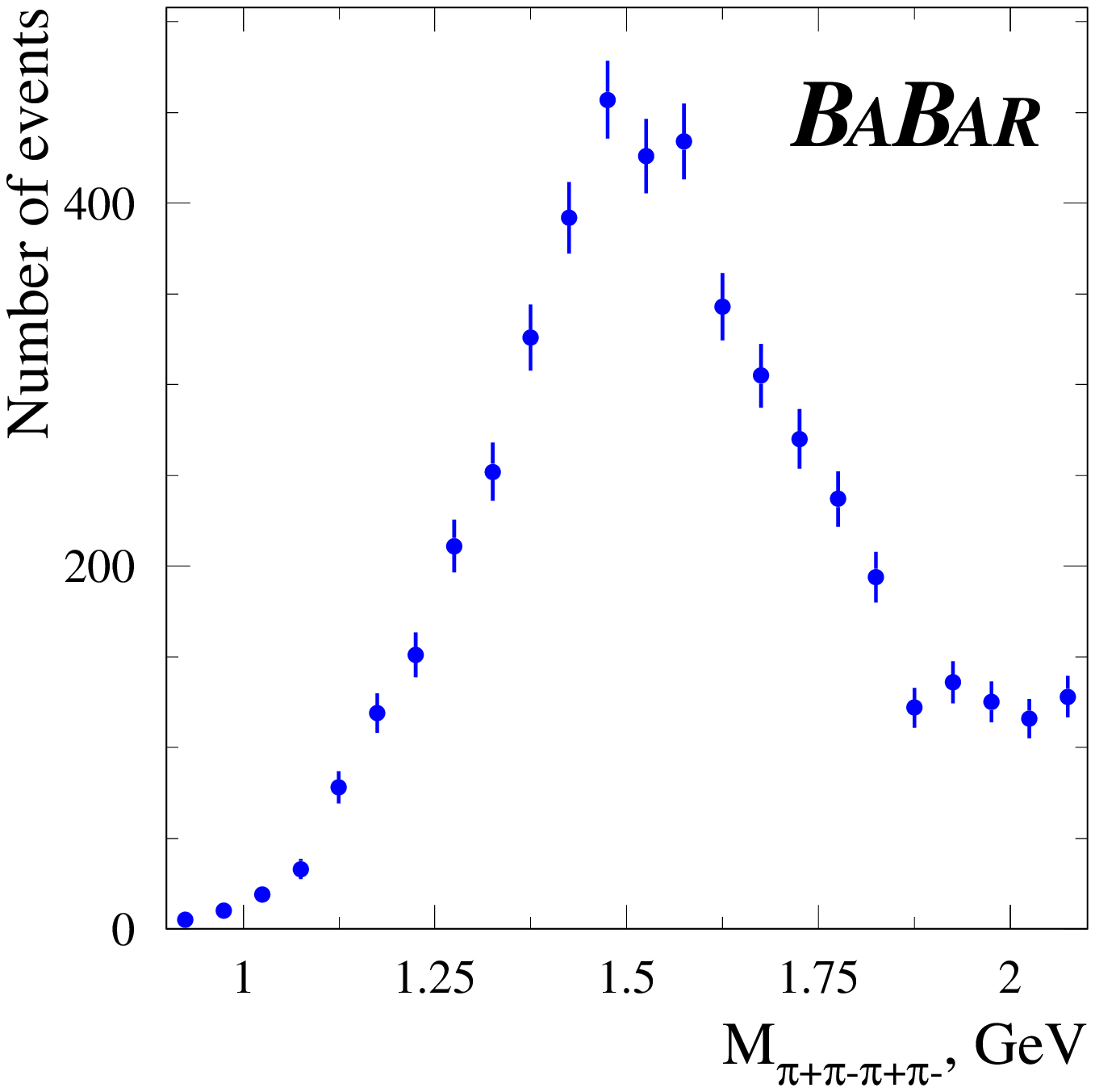}
}
\vspace{-0.3cm}
\caption
{ The c.m. energy dependence of the $\pipi\pipi$ cross section 
 measured in $\ep\en$ experiments (top) compared to the corresponding
uncorrected mass distribution from \babar~ ISR data (bottom).
}
\label{4pi_ee_babar}
\end{figure}
\begin{figure}[tbh]
{
\epsfxsize=9.0cm
\epsfysize=8.0cm
\epsfbox{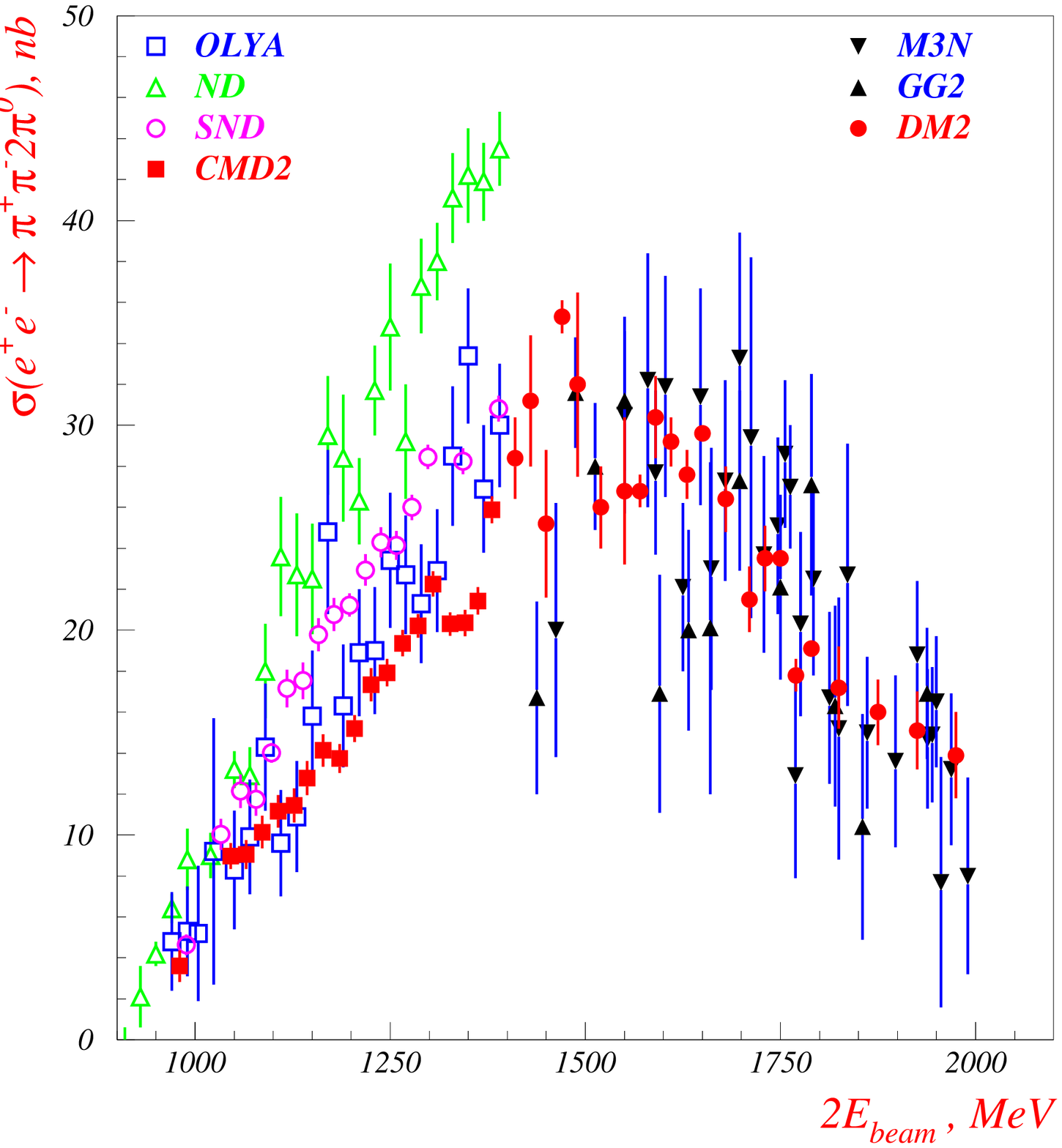}
\epsfxsize=8.5cm
\epsfysize=8.0cm
\epsfbox{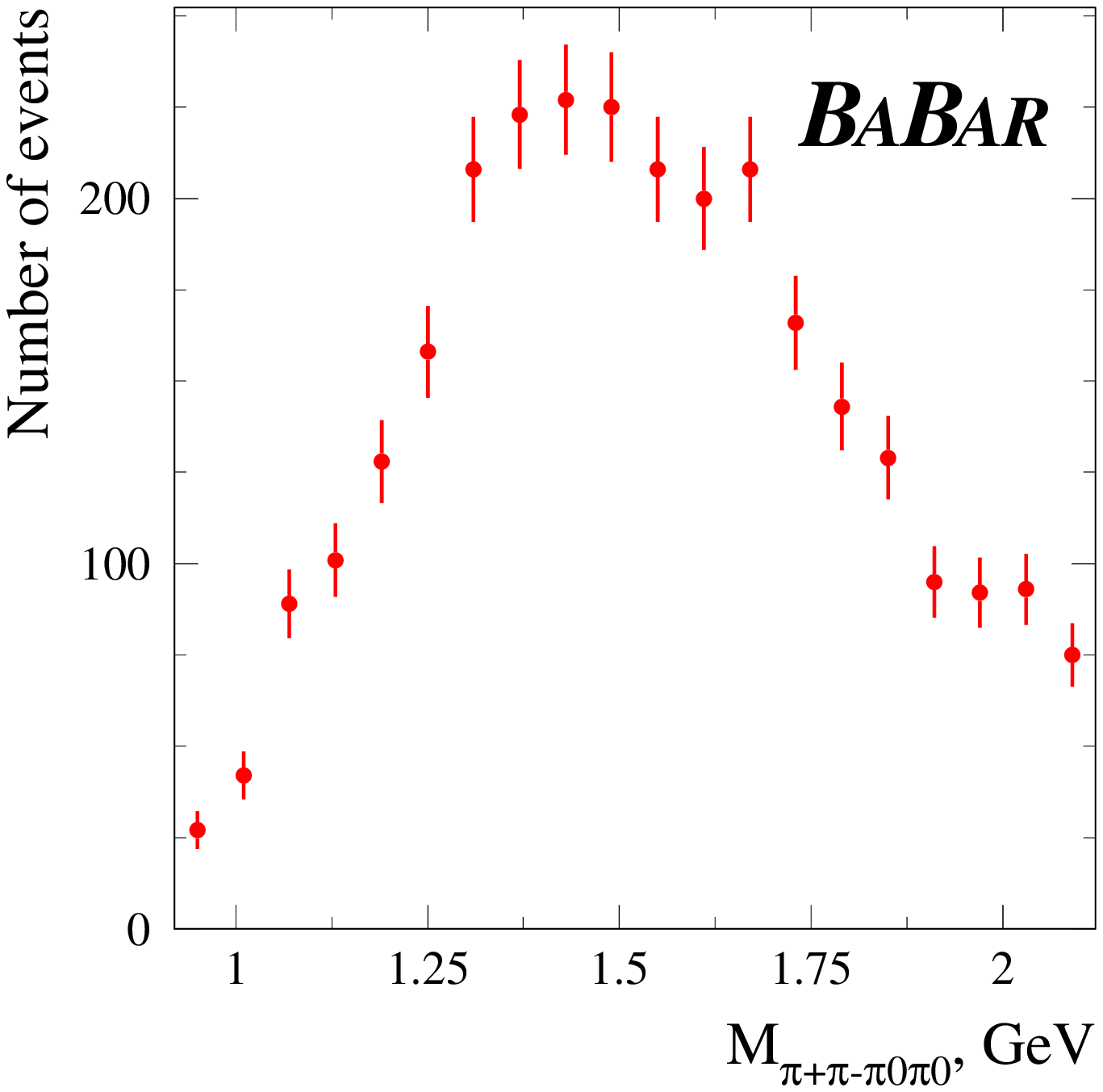}
}
\vspace{-0.9cm}
\caption
{ The c.m. energy dependence of the 
 $\pipi\pi^0\pi^0$ cross section
measured in $\ep\en$ experiments (top) compared to the corresponding
uncorrected mass distribution from \babar~ ISR data (bottom).
}
\label{2pi2pi0_ee_babar}
\end{figure}
\vspace{-0.9cm}

\subsection{$\pi^+\pi^-\pi^0\pi^0$ final state }

This final state is selected by requiring that two charged
tracks and five photons be produced. The most energetic photon is
assumed to be due to ISR, and the mass distribution for all possible
pair combinations of the other four is shown in
fig.~\ref{2pi2pi0_babar} (left). Peaks due to  $\pi^0$ and $\eta$
production are seen. The requirement that two distinct combinations
be in the $\pi^0$ mass range yields the  four-pion
mass distribution shown in the right-hand plot. It is very
similar to that obtained for four charged pions.

The \babar~ data can be compared to existing $\ep\en$ cross section 
measurements for four-pion final states.
 Fig.~\ref{4pi_ee_babar} compares four-charged-pion cross section
measurements from $\ep\en$ colliders (top) to the
corresponding \babar~ ISR data.
A similar comparison is made in fig.~\ref{2pi2pi0_ee_babar} for
$\pi^+\pi^-\pi^0\pi^0$ data. In general, the ISR data and the cross
section data seem to agree in shape, but it is clear that the former
suffer much less from the relative normalization uncertainties
present in the latter.

\begin{figure}[tbh]
\hspace{-0.3cm}
{
\epsfysize=4.5cm
\epsfxsize=4.3cm
\epsfbox{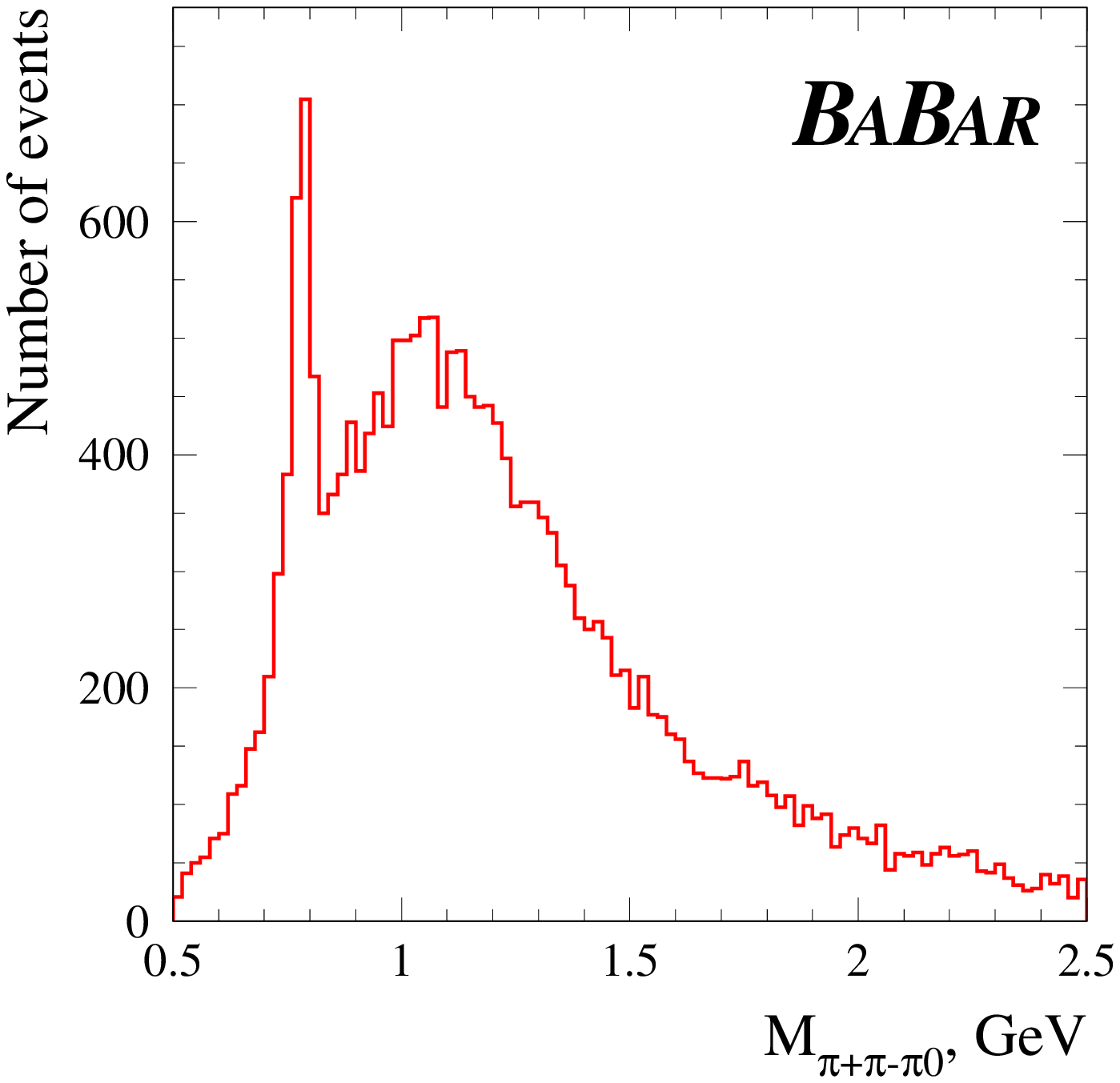}
\hspace{-0.4cm}
\epsfysize=4.5cm
\epsfxsize=4.3cm
\epsfbox{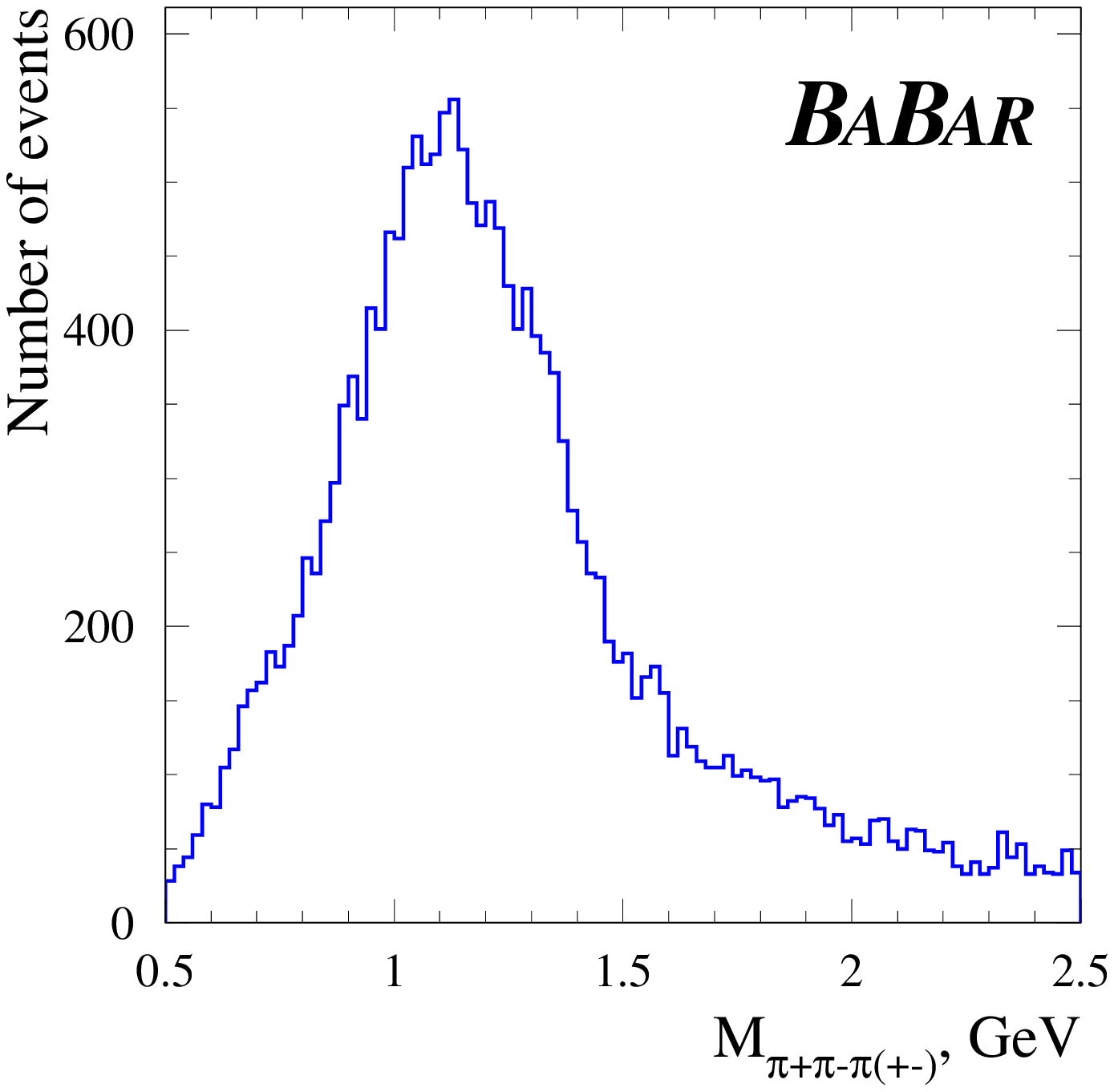}
}
\vspace{-0.9cm}
\caption
{Three pion mass distributions for the
$\pipi\pi^0\pi^0$ and $\pipi\pipi$ final states produced via ISR.
}
\label{3pi_from_4pi}
\end{figure}

\subsection{Combined Analysis}

For a multi-pion final state, it is important to understand the
mass structure present in the contributing multi-pion subsystems,
both from the standpoint of spectroscopy, and for the
understanding of acceptance effects. In this regard, it is very useful
to have information available for different isotopic spin
configurations. This has been demonstrated in a four-pion analysis
performed by the CMD-2 group~\cite{4pi_cmd}, which was recently
confirmed by CLEO using $\tau$ decays~\cite{4pi_tau_cleo}. Both
analyses demonstrate $a_{1}(1260)\pi$ dominance in four-pion production.

 Further confirmation, and in fact more detailed study (because the
phase space is not limited), can be obtained from ISR production of 
four-pion final states. For example,
fig.~\ref{3pi_from_4pi} shows the three-pion
mass distributions resulting from the ISR production of
$\pipi\pipi$ and $\pipi\pi^0\pi^0$ events in the \babar~ data.
It is clear that an
intermediate $\omega\pi^0$ state contributes to
$\pipi\pi^0\pi^0$, and that both final states exhibit a broad peak
which may be due to the $a_{1}(1260)$ resonance.

\section{Higher Multiplicity Final States}

The good resolution and PID characteristics of the \babar~ 
detector permit
ready selection and study of even higher multiplicity final states
produced exclusively (i.e.with no undetected final state particles)
via ISR. For example data samples for the hadronic states
$K^+ K^-\pi^+\pi^-$, $2(\pi^+\pi^-)\pi^0$,
$6\pi(charged)$,  $2(\pi^+\pi^-\pi^0) $ and
$3(\pi^+\pi^-)\pi^0 $ have been selected, and already contain a few
thousand events each.
Normalization to $\mu^+\mu^-\gamma$ will enable cross section
measurements for these processes in the 1-5 GeV c.m. energy range.
 These final states have not been studied
at $\ep\en$ colliders, and so new information on meson spectroscopy
etc. is already contained in the present \babar~ data sample, with
the prospect of a factor of ten increase in statistics over the
next few years.

\section{Summary}

\begin{itemize}

\item{The good resolution and PID capabilities of
the \babar~ detector permit the identification and measurement of a
broad range of final states produced at low effective c.m. energy 
via ISR during data-taking in the $\Upsilon(4S)$ energy region.
}
\item{
Specifically, the observation of low mass $1^{--}$ resonances has
been demonstrated, and it appears that a great deal of additional 
information about meson spectroscopy can be extracted from the data.
}
\item{
The present \babar~ data in the 1.4-3.0 GeV mass range are already
comparable in quality and precision to direct measurements
from the DCI and ADONE machines, and do not suffer from the relative
normalization uncertainties which seem to exist for certain final
states.
}
\item{
If luminosity and efficiency can be understood with
2-3\% accuracy, ISR production with the \babar~ detector should yield
useful measurements of R, the ratio of the hadronic and dimuon
cross section values, in the low-energy regime of
$\ep\en$ collisions.
}
\end{itemize}



\begin{thebibliography}{99}

\bibitem{ivanch} 

M.Benayoun, S.I.Eidelman, V.N.Ivanchenko,
Z.K.Silagadze, ``Spectroscopy at B-factories Using Hard Photon
Emission'',
Modern Phys.Lett. A. Vol.14, No.37(1999)2605.

\bibitem{kuehn} S.Binner et al., Phys.Lett B{\bf459}(1999)279;
J.Kuehn, hep-ph/0101100.



\bibitem{vuco} 

Vuko Brigljevic (LLNL), ``Study of $e^+ e^-
 \to\phi\gamma$ events '',
\babar~ internal documentation, July 14, 2000.

\bibitem{Lou} 

X.C.Lou (UT Dallas), W.Dunwoodie (SLAC),
``Production of the $\psi
(2S)$ via Initial State Radiation at the $\Upsilon (4S)$ Energy'',
\babar~ internal documentation, Aug. 12, 2000.

\bibitem{pdg}D.E.Groom et al., Eur. Phys. J. C{\bf15} (2000)1.

\bibitem{mu_sel_eff} F.Fabozzi et al., ''Muon Identification in the BaBar
Experiment'', \babar~ internal documentation, June 21, 2000.

\bibitem{cmd2_pipi} R.R.Akhmetshin et al., ``Measurement of pion form
factor around $\rho$ resonance with CMD-2 detector'', BINP Preprint
99-10, Novosibirsk, 1999.

\bibitem{om_snd} M.N.Achasov et al. (SND Collaboration),
Phys.Lett. B{\bf462}(1999)365.


\bibitem{4pi_cmd}  R.R.Akhmetshin et al.(CMD-2 Collaboration),
Phys.Lett. B{\bf466}(1999)392.

\bibitem{4pi_tau_cleo}A.J.Weinstein et al.(CLEO Collaboration),
''Semihadronic $\tau$ decays at CLEO'',
Nucl.Phys.Proc.Suppl.{\bf98}(2001)261.

\bibitem{ompipi_cmd2} R.R.Akhmetshin et al.(CMD-2 Collaboration),
Phys.Lett. B{\bf489}(2000)125.




\end{thebibliography}
\end{document}